\documentclass[aps,showkeys,superscriptaddress,preprint,tightenlines]{revtex4-2}
\usepackage{hyperref}
\usepackage{times}
\usepackage{amsmath}
\usepackage{graphicx}
\usepackage{epsfig}
\usepackage{dcolumn}
\usepackage{bm}
\usepackage{color}
\usepackage{longtable}
\usepackage{array}
\usepackage{multirow}
\usepackage{setspace}
\usepackage[mathlines]{lineno}
\usepackage{epsfig,graphics,subfigure,psfrag}
\newcommand{\Ecm}{E_{\rm cm}}
\newcommand{\xyz}{\rm XYZ}
\newcommand{\lum}{{\cal L}}

\newcommand{\jpsi}{J/\psi}
\newcommand{\EE}{e^+e^-}
\newcommand{\MM}{\mu^+\mu^-}

\newcommand{\beq}{\begin{equation}}
\newcommand{\eeq}{\end{equation}}
\newcommand{\bitm}{\begin{itemize}}
\newcommand{\eitm}{\end{itemize}}

\begin{document}

\title{\boldmath Measurements of the center-of-mass energies
of $\EE$ collisions at BESIII}


\author{
\author{Author list}
		\begin{small}
		\begin{center}
	M.~Ablikim$^{1}$, M.~N.~Achasov$^{10,c}$, P.~Adlarson$^{67}$, S. ~Ahmed$^{15}$, M.~Albrecht$^{4}$, R.~Aliberti$^{28}$, A.~Amoroso$^{66A,66C}$, M.~R.~An$^{32}$, Q.~An$^{49,63}$, X.~H.~Bai$^{57}$
% ...
\\
\vspace{0.2cm}
(BESIII Collaboration)\\
\vspace{0.2cm} {\it
$^{1}$ Institute of High Energy Physics, Beijing 100049, People's Republic of China\\
% ...
$^{4}$ Bochum Ruhr-University, D-44780 Bochum, Germany\\
% ...
$^{10}$ G.I. Budker Institute of Nuclear Physics SB RAS (BINP), Novosibirsk 630090, Russia\\
% ...
$^{15}$ Helmholtz Institute Mainz, Johann-Joachim-Becher-Weg 45, D-55099 Mainz, Germany\\
% ...
$^{28}$ Johannes Gutenberg University of Mainz, Johann-Joachim-Becher-Weg 45, D-55099 Mainz, Germany\\
% ...
$^{32}$ Liaoning Normal University, Dalian 116029, People's Republic of China\\
% ...
$^{49}$ State Key Laboratory of Particle Detection and Electronics, Beijing 100049, Hefei 230026, People's Republic of China\\
% ...
$^{57}$ University of Jinan, Jinan 250022, People's Republic of China\\
% ...
$^{63}$ University of Science and Technology of China, Hefei 230026, People's Republic of China\\
% ...
$^{66}$ (A)University of Turin, I-10125, Turin, Italy; (B)University of Eastern Piedmont, I-15121, Alessandria, Italy; (C)INFN, I-10125, Turin, Italy\\
$^{67}$ Uppsala University, Box 516, SE-75120 Uppsala, Sweden\\
% ...
$^{c}$ Also at the Novosibirsk State University, Novosibirsk, 630090, Russia\\
% ...
		}\end{center}
		\vspace{0.6cm}
		\end{small}
}

\date{\today}

\begin{abstract}

During the 2016-17 and 2018-19 running periods,
the BESIII experiment collected 7.5~fb$^{-1}$ of $e^+e^-$ collision data at
center-of-mass energies ranging from 4.13 to 4.44~GeV.
These data samples are primarily used for the study of excited charmonium and charmoniumlike states.
By analyzing the di-muon process $\EE \to (\gamma_{\rm ISR/FSR}) \MM$,  we measure the center-of-mass energies of the data samples with a precision of 0.6~MeV.
Through a run-by-run study, we find that the center-of-mass energies
were stable throughout most of the data-taking period.
\end{abstract}


\maketitle

\section{Introduction}

The BESIII experiment~\cite{bes3} was designed to study physics in
the $\tau$-charm energy region (2.0 -- 4.9 GeV)~\cite{yellowbook}
with $\EE$ annihilation produced by the BEPCII storage
ring~\cite{bepc2}. Since it started running in 2008, a variety of data
samples have been collected at different center-of-mass~(CM) energies
for the study of light hadron spectroscopy, charmonium and
charmoniumlike states (also called $\xyz$ states), charm physics,
$\tau$ physics, various QCD-related studies, and the search for
new physics beyond the standard model~\cite{whitepaper}.

The Beam Energy Measurement System~(BEMS)~\cite{BEMS} was designed to precisely measure BESIII CM energies~($\Ecm$) using a method based on
Compton back-scattered photons. However, its capability at high energy~($\Ecm$ above 4~GeV) is degraded by its detection efficiency and limited calibration
sources for high-energy gamma rays.
Therefore, an alternative algorithm was
developed to measure the $\Ecm$ for data samples above 4~GeV. This method uses the
well-understood QED process $\EE\to (\gamma_{\rm ISR/FSR}) \MM$
(the di-muon process), where $\gamma_{\rm
ISR/FSR}$ is a radiative photon due to initial state radiation
(ISR) and/or final state radiation (FSR).  Using this method, a precision of 0.8~MeV was previously achieved for the
data taken from 2011 to 2014~\cite{ecm_XYZ}.

In this paper, we present the $E_{\rm cm}$ measurement for the
$\xyz$ data samples taken at BESIII from 2017 to 2019. The method
used in Ref.~\cite{ecm_XYZ} is followed, but the precision of
the momentum calibration is improved, and the $\Ecm$ is measured with
an uncertainty of 0.6~MeV.

Using the selected di-muon events, $\EE\to (\gamma_{\rm ISR/FSR})
\MM$, we determine $\Ecm$ with
\begin{equation}
\label{Ecmscal}
    \Ecm = (M_{\rm p}(\MM) + \Delta M_{\rm ISR/FSR} + \Delta M_{\rm cal})\times c^{2},
\end{equation}
where $M_{\rm p}(\MM)$ is the peak position of the $\MM$ invariant
mass of selected di-muon events; $\Delta M_{\rm ISR/FSR}$ is the
mass shift due to the emission of ISR or FSR photons, estimated
from Monte Carlo (MC) simulation of the di-muon process by turning on
and off the ISR/FSR processes in MC generation; and $\Delta M_{\rm cal}$ is
the correction introduced by the momentum calibration of the $\MM$
tracks, obtained from an analysis of the process $\EE\to \gamma_{\rm ISR}
J/\psi$.

\section{The BESIII detector and data sets}

The BESIII detector is described in detail in Ref.~\cite{bes3}.
The cylindrical core of the detector covers 93\% of the full solid
angle and consists of a helium-based multilayer drift
chamber~(MDC), a plastic scintillator time-of-flight system~(TOF),
and a CsI(Tl) electromagnetic calorimeter~(EMC), which are all
enclosed in a superconducting solenoidal magnet providing a 1.0~T
magnetic field. The solenoid is supported by an octagonal
flux-return yoke with resistive plate counter muon identification
modules interleaved with steel. The charged-particle momentum
resolution at $1~{\rm GeV}/c$ is $0.5\%$, and the $dE/dx$
resolution is $6\%$ for electrons from Bhabha scattering. The EMC
measures photon energies with a resolution of $2.5\%$ ($5\%$) at
$1$~GeV in the barrel (end cap) region. The time resolution in the TOF barrel region is 68 ps
while that in the end cap region is 60 ps~\cite{etof}.


The data samples analyzed in this work are listed in Table~\ref{ResultXYZ}.
They include 16 different
CM energies from 4.13 to 4.44~GeV and were collected in two running years: from
December 2016 to May 2017 (labelled as ``2017XYZ" hereafter, the integrated luminosities are measured using the Bhabha events in Ref.~\cite{yangyf}.); and from
February 2019 to June 2019 (labelled as ``2019XYZ" hereafter, the integrated luminosities are estimated by using online monitoring information). The column "Sample" shows the
nominal CM energy in MeV used during online data taking. The true
CM energy is usually within a few MeV of the nominal value.
Run numbers are used to divde the data into subsamples. Other columns, such as $\mathcal{L}$ ($\rm{pb^{-1}}$), will be illustrated below.

\begin{table}[htbp]
 \centering
\caption{Summary of the data samples, including run numbers,
integrated luminosity $\lum$~\cite{yangyf}, the measured $\jpsi$
mass after FSR correction $M^{\rm cor}(\jpsi)$ (in MeV/$c^2$),
$M_{\rm p}(\MM)$ (in MeV/$c^2$), and $\Ecm$. Superscripts
represent data from different periods: ``1" stands for 2017XYZ
data, and ``2" stands for 2019XYZ data. The first uncertainties
are statistical, and the second systematic.}\label{ResultXYZ}
\begin{tabular}{cclccc}
\hline
\hline
  Sample & Run Number & $\mathcal{L}$ ($\rm{pb^{-1}}$) & $M^{\rm cor}(J/\psi) $ & $M_{\rm p}(\mu^{+}\mu^{-})$ & $E_{\rm cm}$ (MeV) \\
\hline
  4130$^2$ & 59163-59573   & 400  &    $3100.55\pm0.30$    &    $4130.23\pm0.05$  &  $4128.78\pm0.05\pm0.36$ \\
  4160$^2$ & 59574-59896   & 400  &    $3100.18\pm0.29$    &    $4158.89\pm0.05$  &  $4157.83\pm0.05\pm0.34$ \\
  4190$^1$ & 47543-48170   & $526.70\pm2.16$  &    $3097.89\pm0.28$    &    $4187.90\pm0.05$  &  $4189.12\pm0.05\pm0.34$ \\
  4200$^1$ & 48172-48713   & $526.60\pm2.05$  &    $3098.17\pm0.27$    &    $4198.20\pm0.05$  &  $4199.15\pm0.05\pm0.34$ \\
  4210$^1$ & 48714-49239   & $517.10\pm1.81$  &    $3097.41\pm0.29$    &    $4207.67\pm0.06$  &  $4209.39\pm0.06\pm0.34$ \\
  4220$^1$ & 49270-49787   & $514.60\pm1.80$  &    $3097.51\pm0.26$    &    $4217.31\pm0.05$  &  $4218.93\pm0.06\pm0.32$ \\
  4237$^1$ & 49788-50254   & $530.30\pm2.39$  &    $3097.36\pm0.24$    &    $4233.99\pm0.04$  &  $4235.77\pm0.04\pm0.30$ \\
  4246$^1$ & 50255-50793   & $538.10\pm2.69$  &    $3097.35\pm0.24$    &    $4242.18\pm0.04$  &  $4243.97\pm0.04\pm0.30$ \\
  4270$^1$ & 50796-51302   & $531.10\pm3.13$  &    $3098.09\pm0.26$    &    $4265.74\pm0.04$  &  $4266.81\pm0.04\pm0.32$ \\
  4280$^1$ & 51305-51498   & $175.70\pm0.97$  &    $3097.55\pm0.48$    &    $4277.73\pm0.04$  &  $4277.78\pm0.11\pm0.52$ \\
  4290$^2$ & 59902-60363   & 500  &    $3100.07\pm0.28$    &    $4289.33\pm0.06$  &  $4288.43\pm0.06\pm0.34$ \\
  4315$^2$ & 60364-60805   & 500  &    $3099.97\pm0.30$    &    $4313.46\pm0.06$  &  $4312.68\pm0.06\pm0.35$ \\
  4340$^2$ & 60808-61242   & 500  &    $3099.71\pm0.29$    &    $4338.45\pm0.06$  &  $4337.93\pm0.06\pm0.35$ \\
  4380$^2$ & 61249-61762   & 500  &    $3099.68\pm0.30$    &    $4378.35\pm0.06$  &  $4377.88\pm0.06\pm0.35$ \\
  4400$^2$ & 61763-62285   & 500  &    $3100.61\pm0.31$    &    $4398.21\pm0.06$  &  $4396.83\pm0.06\pm0.36$ \\
  4440$^2$ & 62286-62823   & 570  &    $3099.73\pm0.29$    &    $4437.59\pm0.06$  &  $4437.10\pm0.06\pm0.35$ \\
\hline
\hline
\end{tabular}
\end{table}

A {\sc geant4}~\cite{GEANT4} based detector simulation package is
developed to model the detector response for MC events. In our
analysis, the di-muon sample is generated with {\sc
babayaga3.5}~\cite{babayaga}, and the $\EE\to \gamma_{\rm
ISR}J/\psi$ sample is generated with {\sc kkmc}~\cite{kkmc}. One
million events are generated for each process at each CM energy.

\section{\boldmath Event selection and measurement of $M_{p}(\MM)$} \label{MeasEcms}

The di-muon process $\EE\to (\gamma_{\rm ISR/FSR})\MM$ is selected
by requiring two oppositely charged tracks in the detector, each
positively identified as a muon. Both charged tracks are reconstructed from hits in the MDC within the polar angle range $|\cos\theta|<0.8$
and their extrapolation to the interaction point (IP) is required to be within 10 cm along the beam direction and within 1 cm in the plane perpendicular to the beam. The energy deposition in the EMC for each charged track is required to be less than
0.4~GeV to suppress backgrounds from radiative Bhabha events.

The sample after these selections includes di-muon events with no
photon emission or with very low-energy radiative photons, ISR
$\jpsi$ with $\jpsi\to \MM$, and ISR $\MM$ events with a smooth
$\MM$ invariant mass ($M(\MM)$) distribution. The events in the $\jpsi$ mass region are used for track momentum calibration
and those with high invariant mass are used to measure the $\Ecm$
after the additional selections criteria are applied as shown below.

To suppress di-muon events with high energy radiative photons, a requirement on the
cosine of the opening angle between the two tracks, $\cos\theta_{\MM} < -0.9997$ is applied. To further remove cosmic ray
events, the TOF time difference between the two tracks is required
to be $|\Delta t| < 2$~ns. The background contribution after the
above selection criteria is less than 0.1\% compared with signal
and is therefore neglected in the following analysis.

\begin{figure}[htbp]
  \centering
  \includegraphics[width=0.6\textwidth, height=0.4\textwidth]{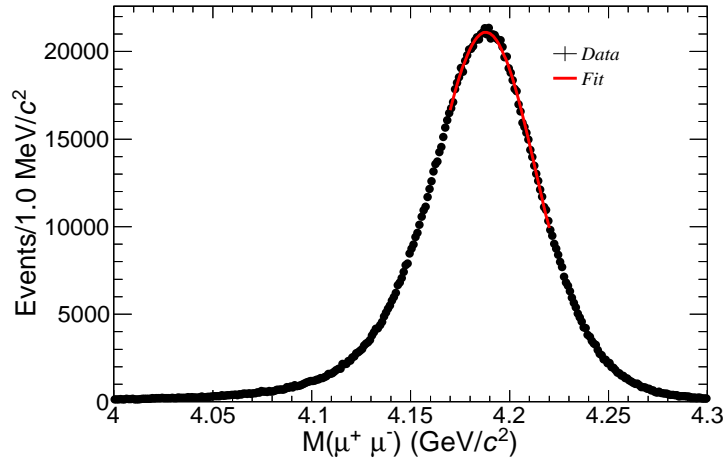}
\caption{The $\MM$ invariant mass distribution and the fit result
of the 4190 sample. Dots with error bars are data, and the red
solid curve is the fit. } \label{Fitmethod}
\end{figure}

The $M(\MM)$ distribution for the 4190 data sample is shown in
Fig.~\ref{Fitmethod} as an example.  The distributions of the other
samples are very similar. The distribution is a Gaussian due to
the momentum resolution of the $\MM$ but is distorted by ISR and
FSR effects which produce a tail on the left side of the peak. The
central part of the distribution can be approximated with a
Gaussian function. We measure the peak position of the
distribution ($M_{\rm p}(\MM)$) by fitting it with a Gaussian
function in a range of $(-1\sigma,~+1.5\sigma)$ around the peak,
where $\sigma$ is the standard deviation of the Gaussian. If the
goodness of the fit, $\chi^2/ndf>2.0$ ($ndf$ is the number of
degrees of freedom of the fit), we slightly reduce the fit
range until $\chi^2/ndf<2.0$ to guarantee a good fit quality. The
fit result for the 4190 data sample is shown in
Fig.~\ref{Fitmethod}. The values of $M_{\rm p}(\MM)$ for the other data samples
are obtained in a similar way and are shown in
Table~\ref{ResultXYZ}.

\begin{figure}[htbp]
  \centering
  \includegraphics[width=0.22\textwidth]{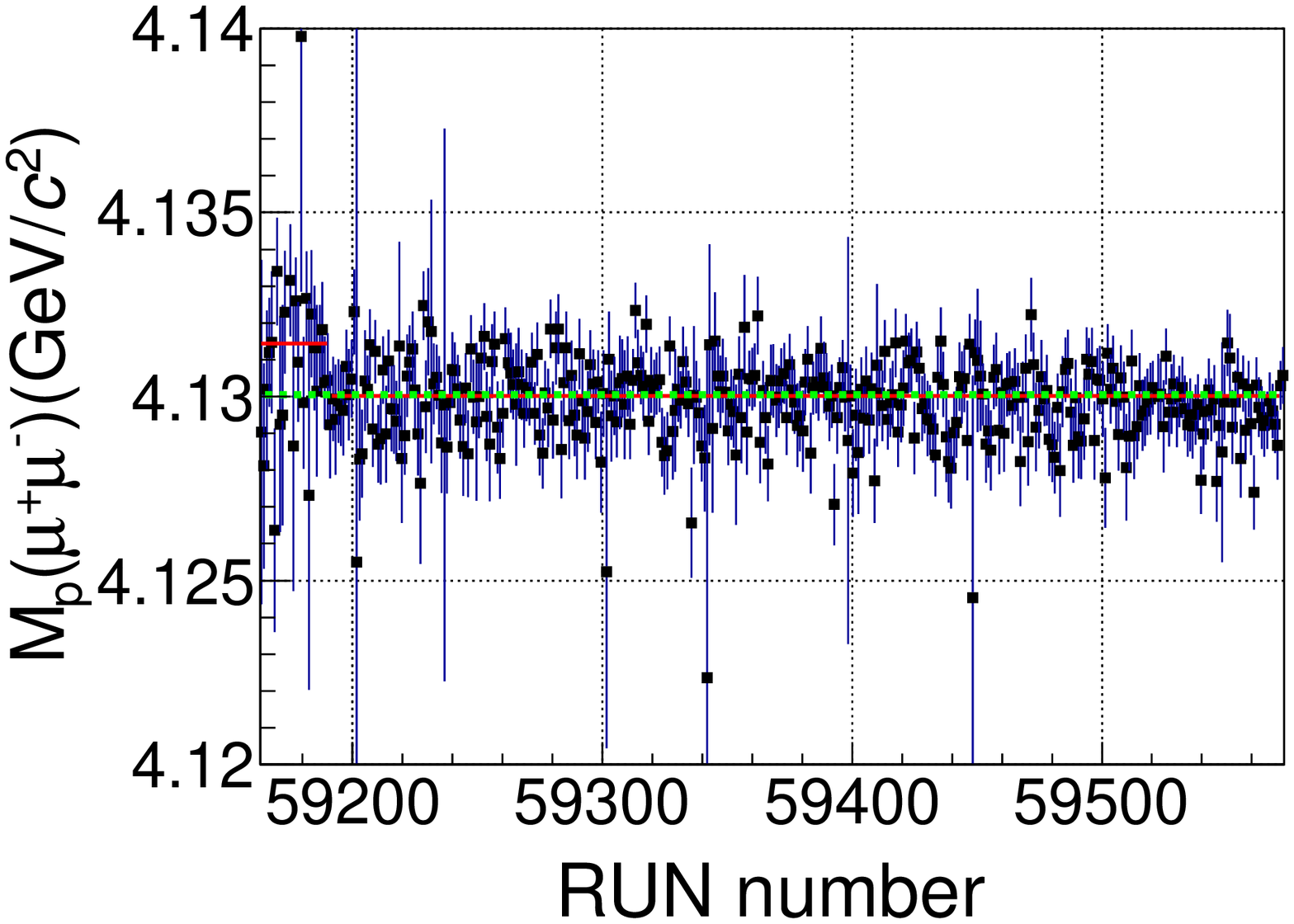}
  \includegraphics[width=0.22\textwidth]{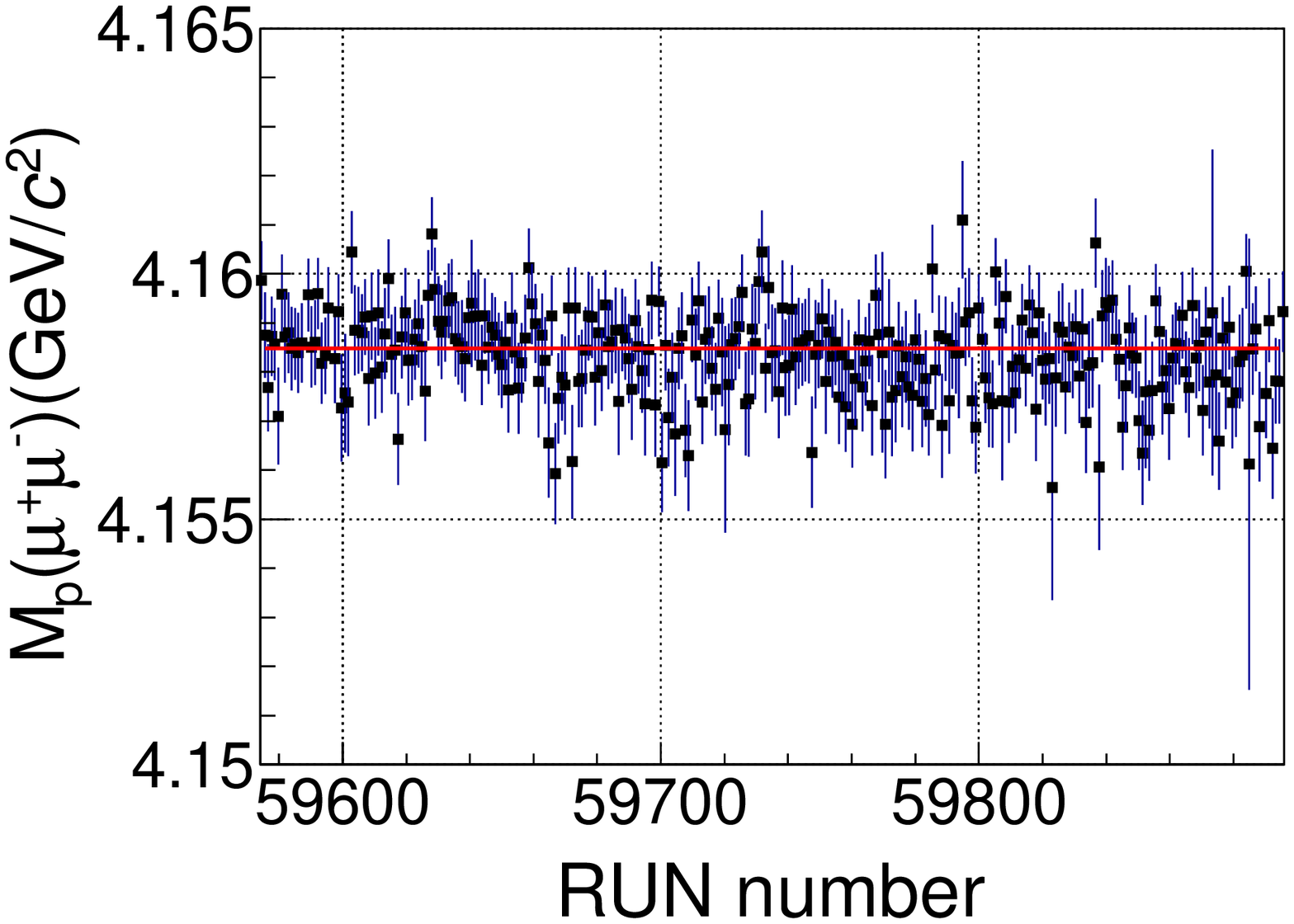}
  \includegraphics[width=0.22\textwidth]{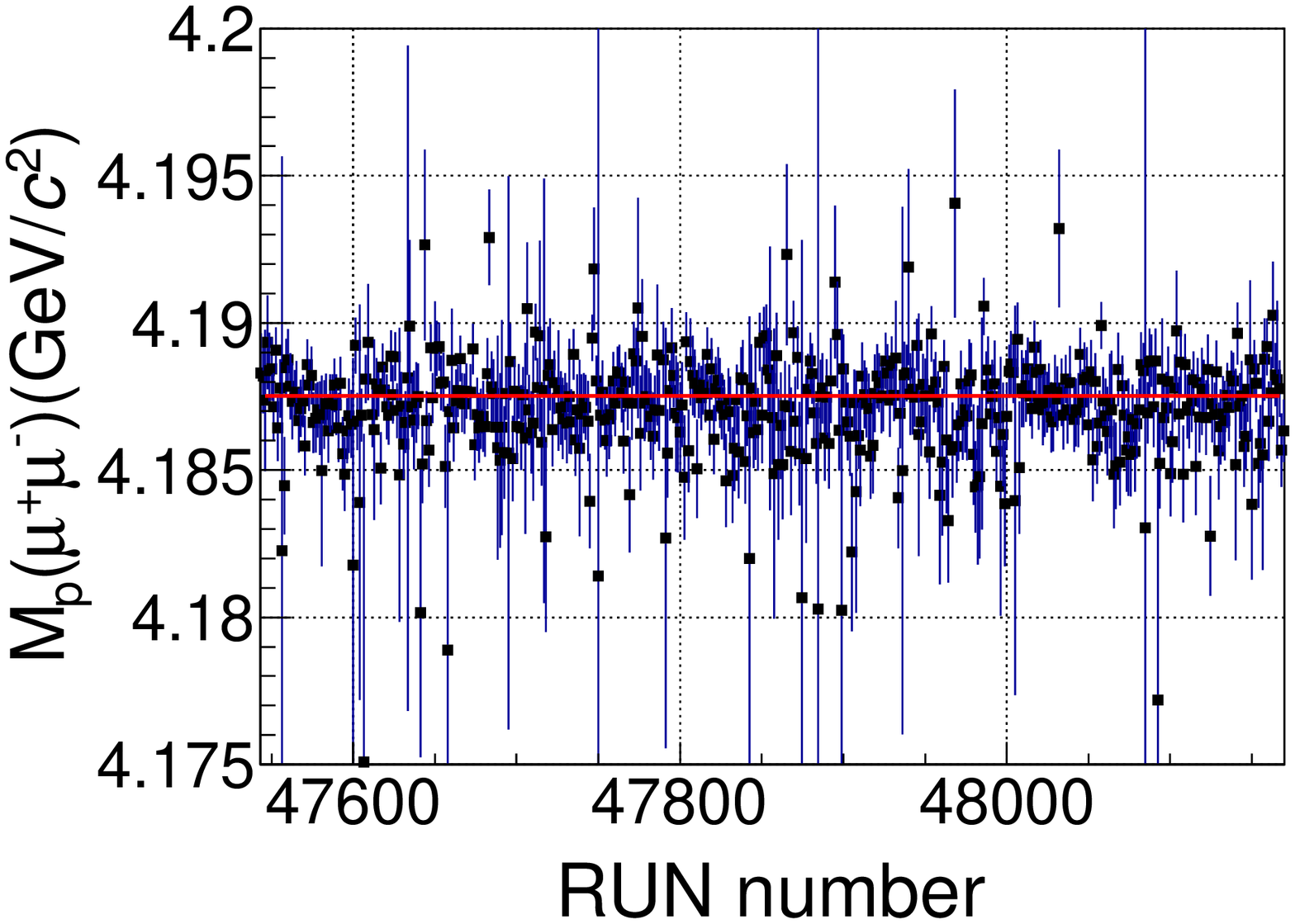}
  \includegraphics[width=0.22\textwidth]{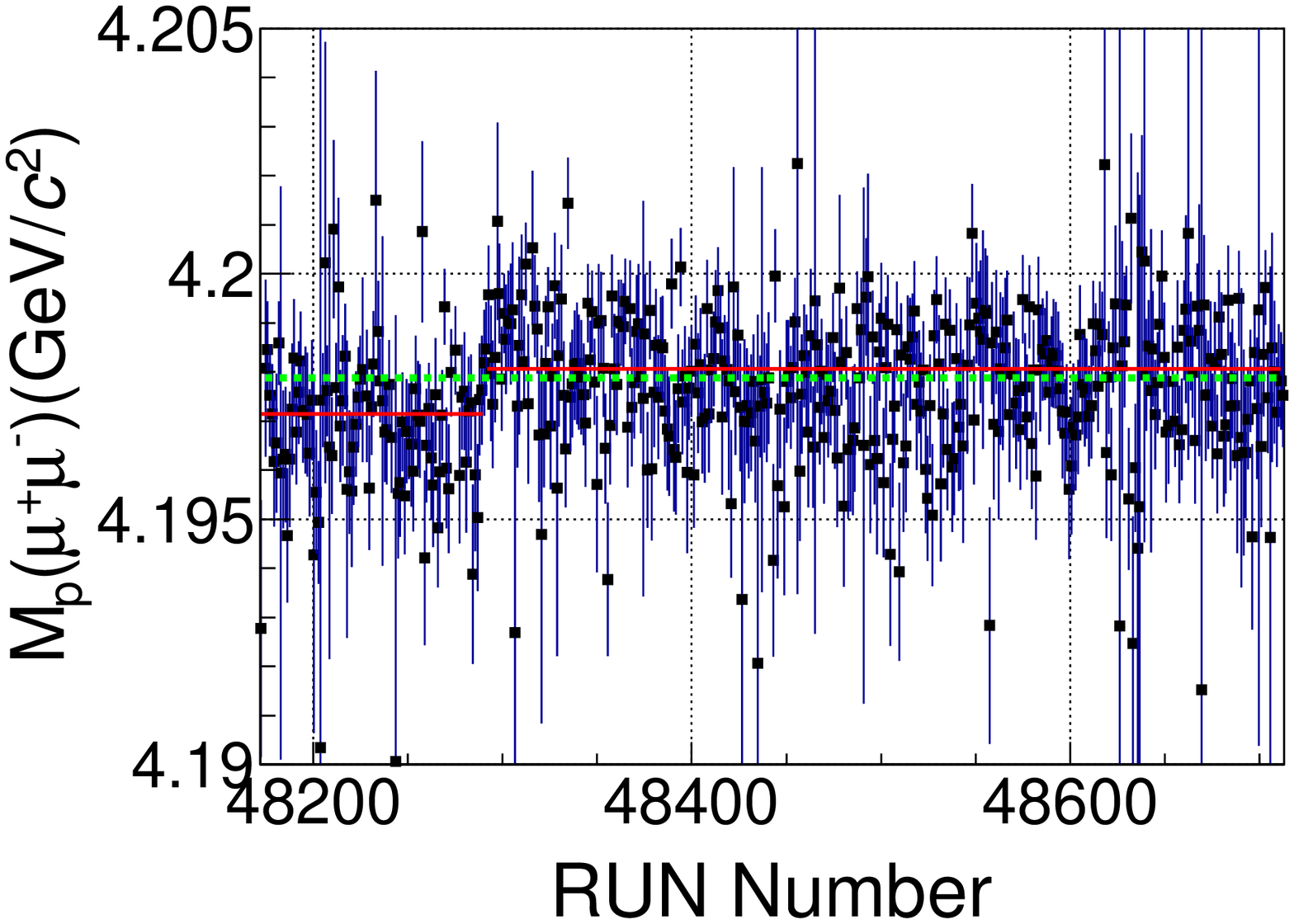}
  \includegraphics[width=0.22\textwidth]{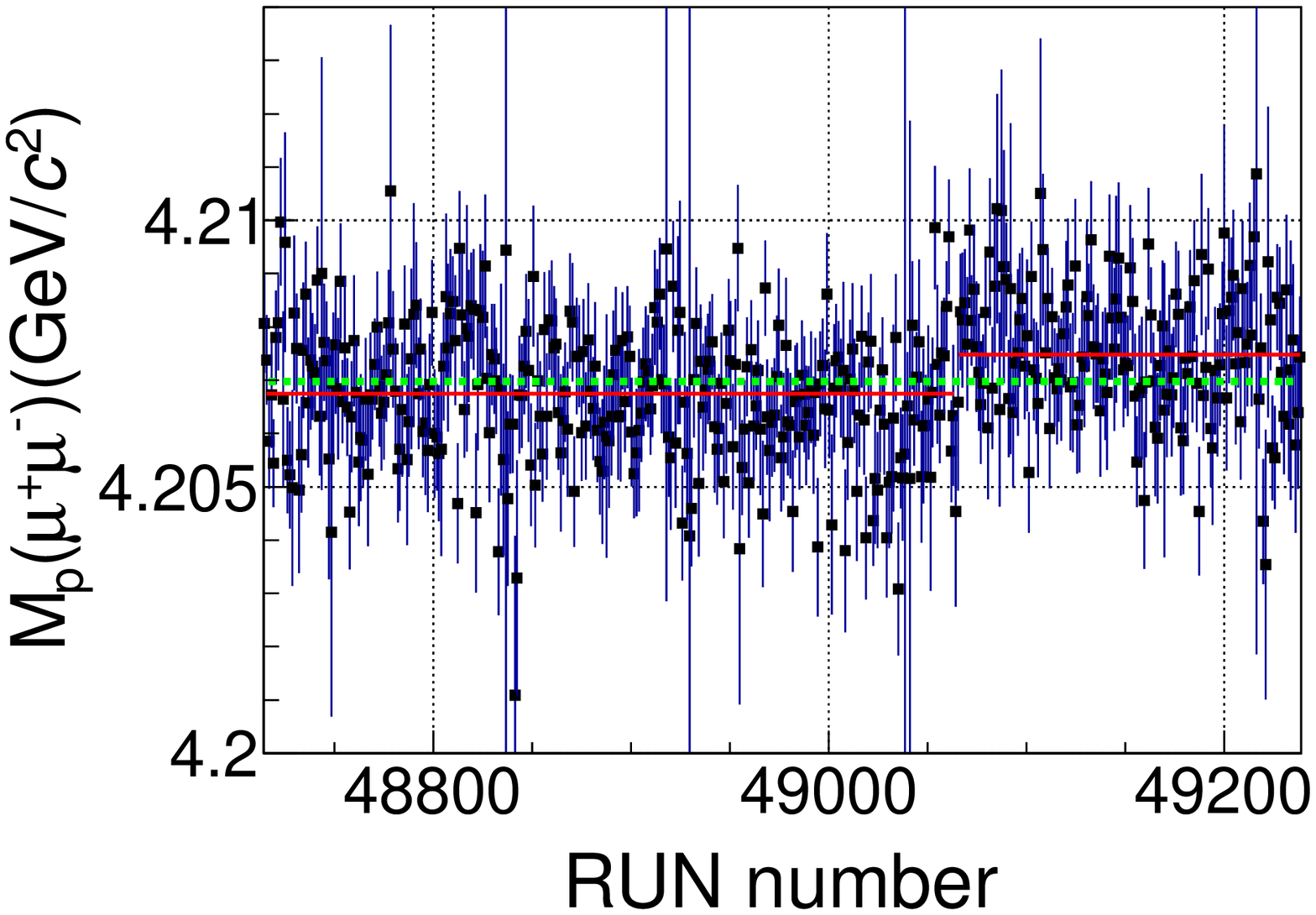}
  \includegraphics[width=0.22\textwidth]{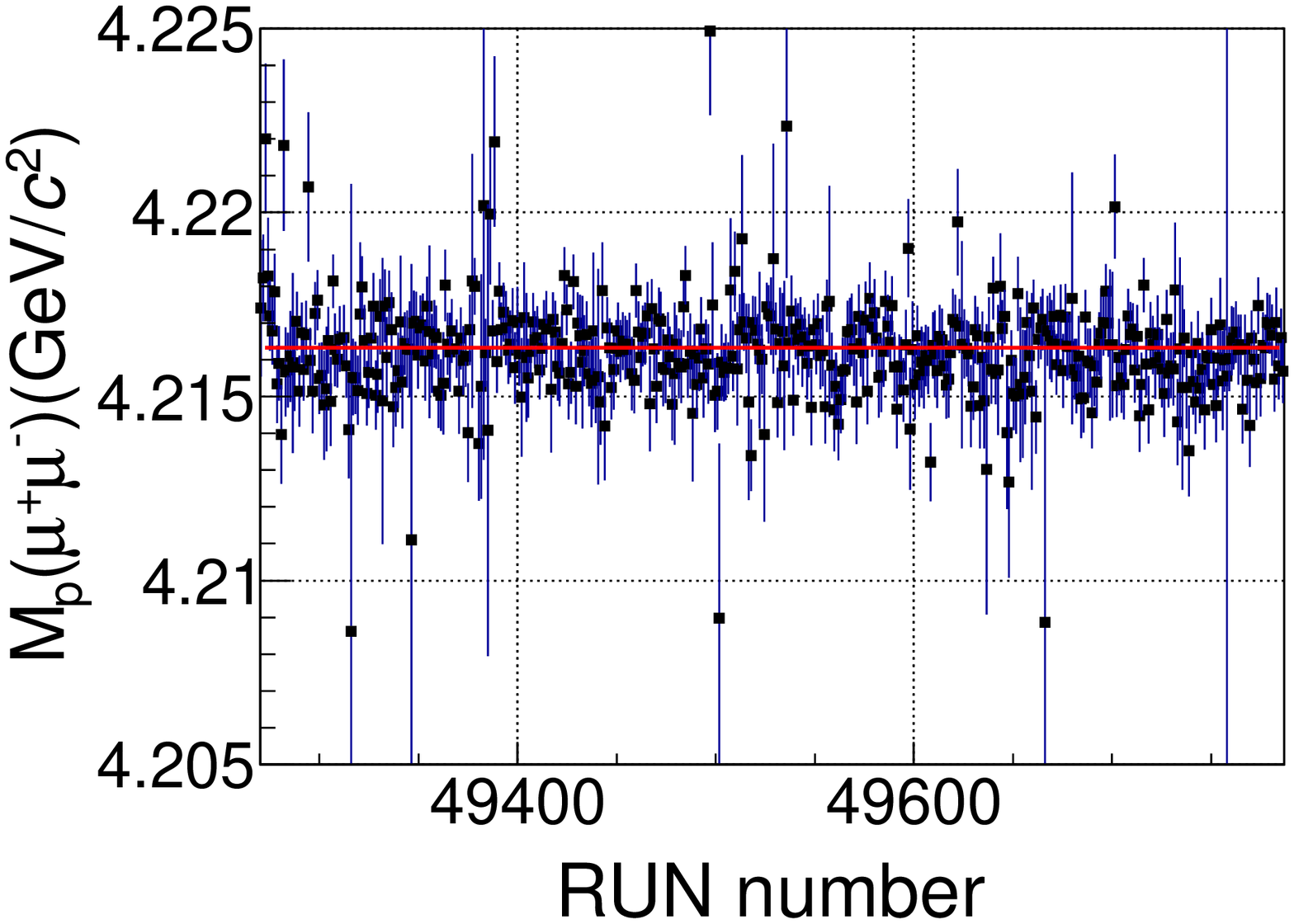}
  \includegraphics[width=0.22\textwidth]{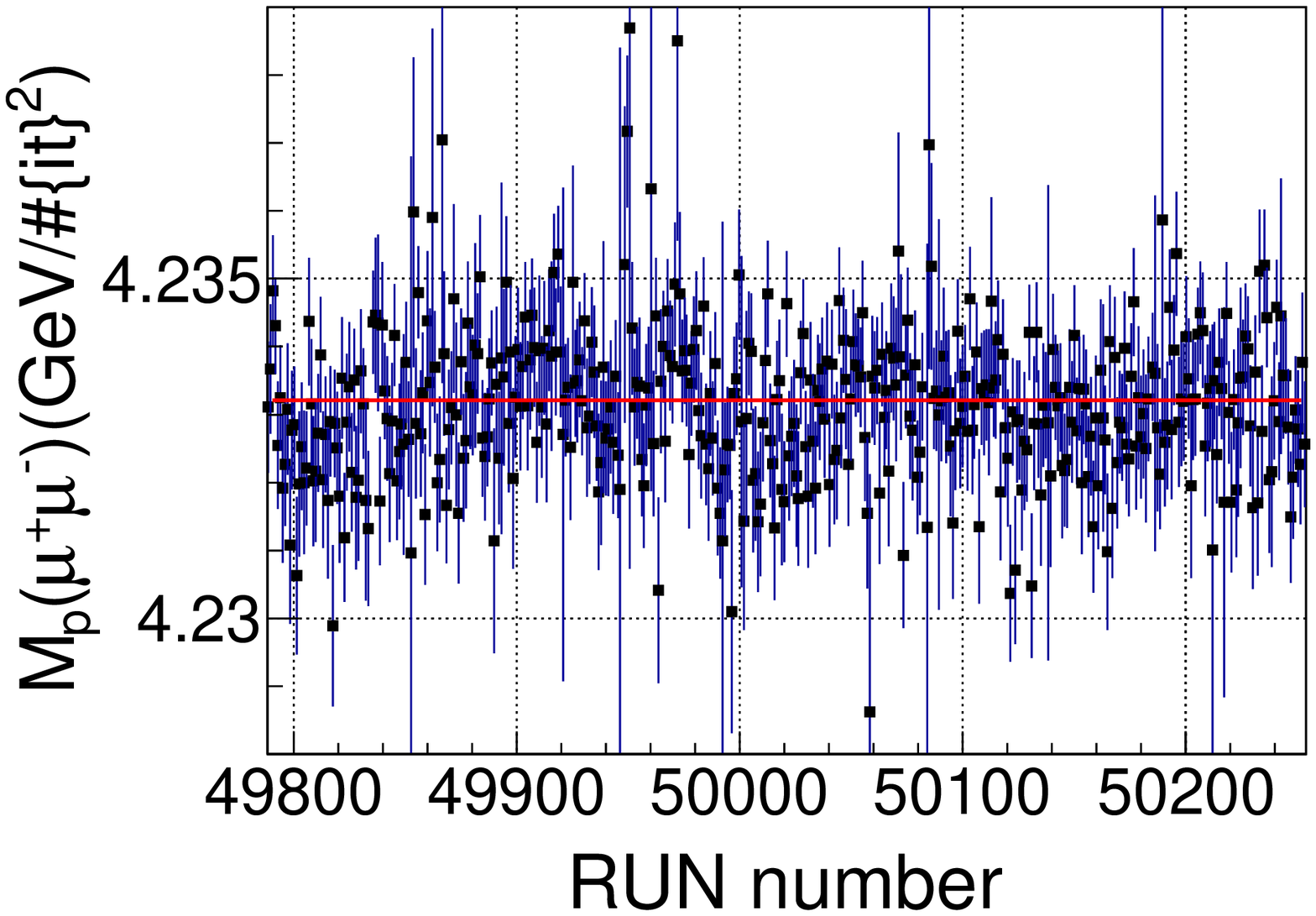}
  \includegraphics[width=0.22\textwidth]{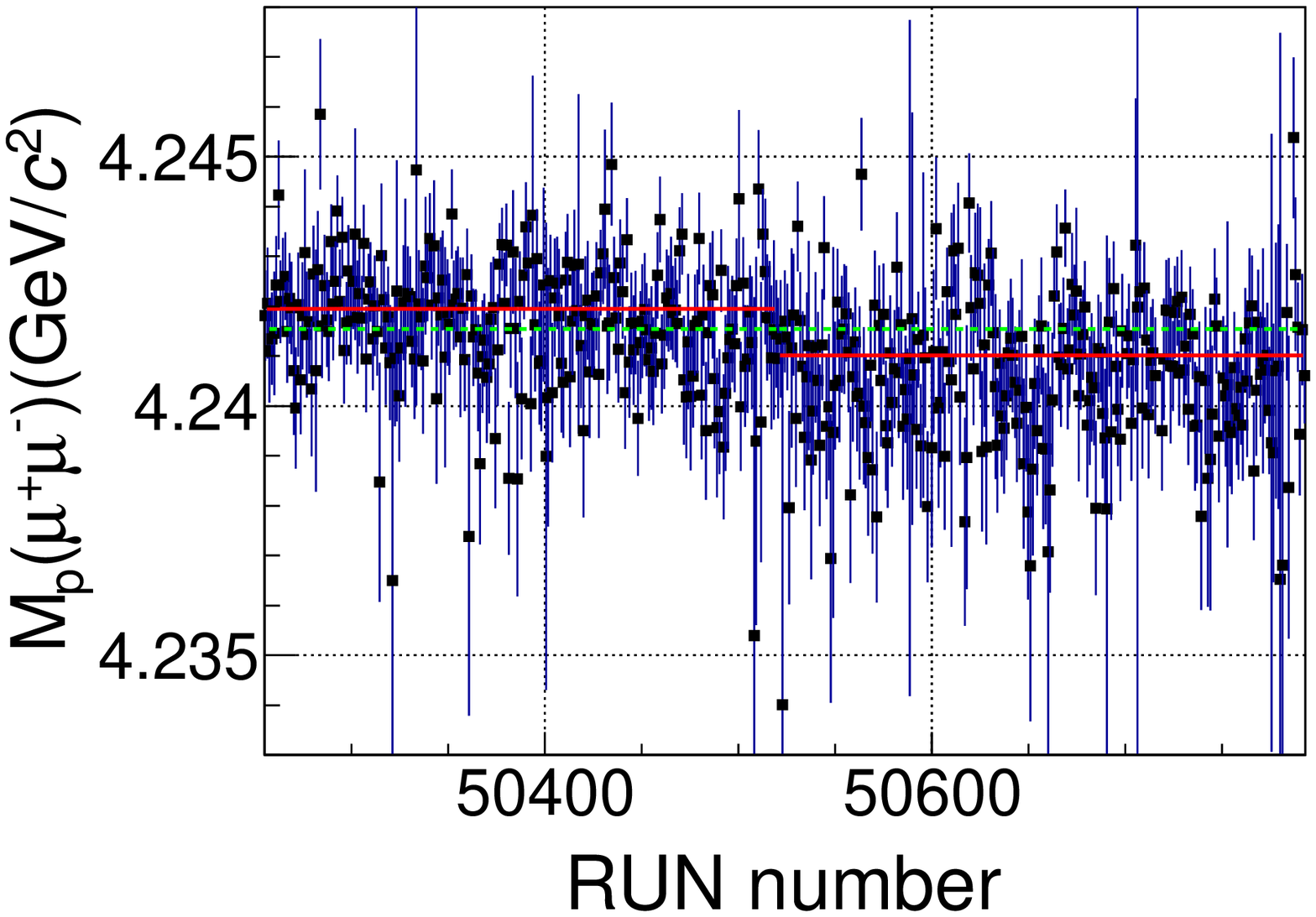}
  \includegraphics[width=0.22\textwidth]{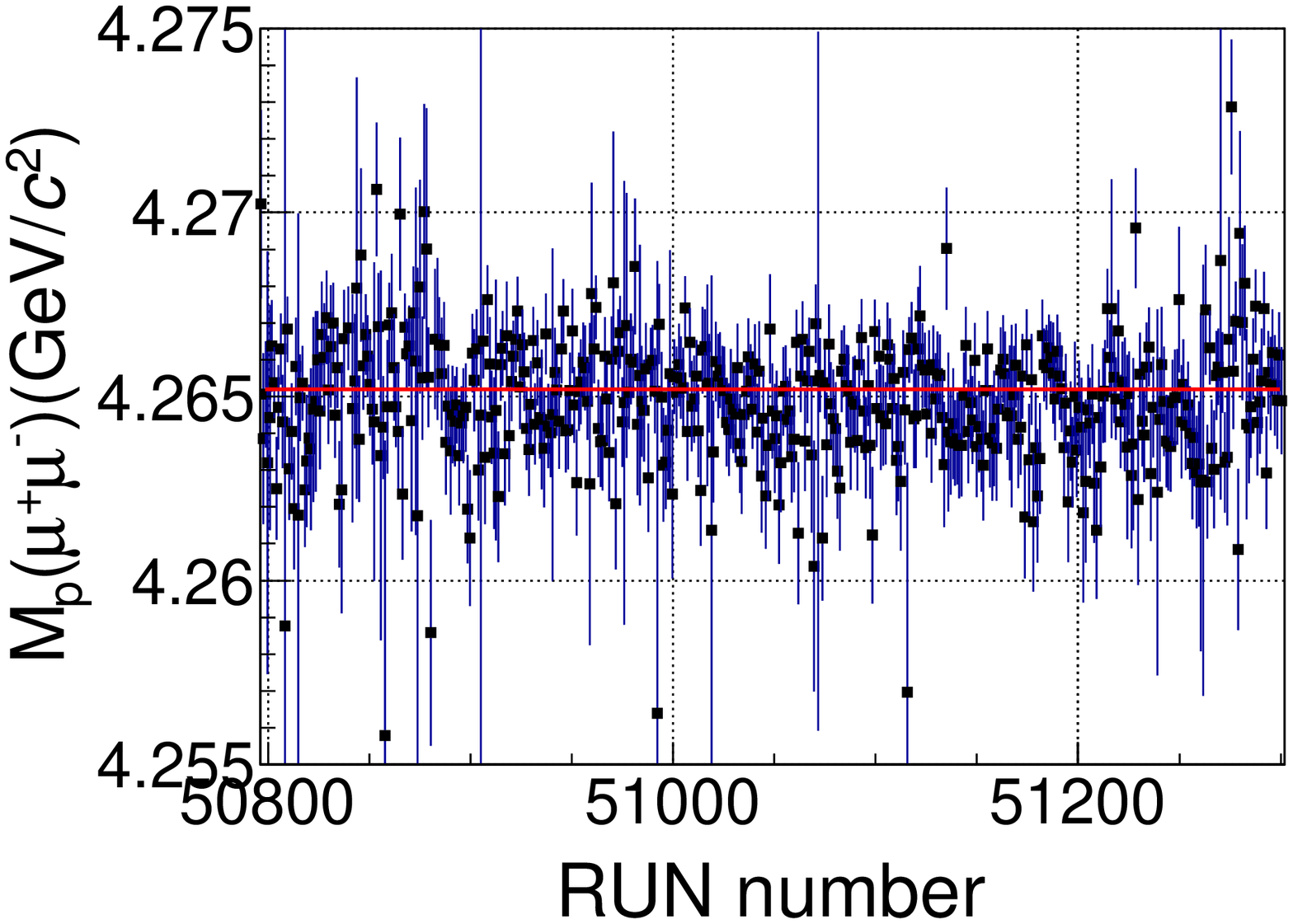}
  \includegraphics[width=0.22\textwidth]{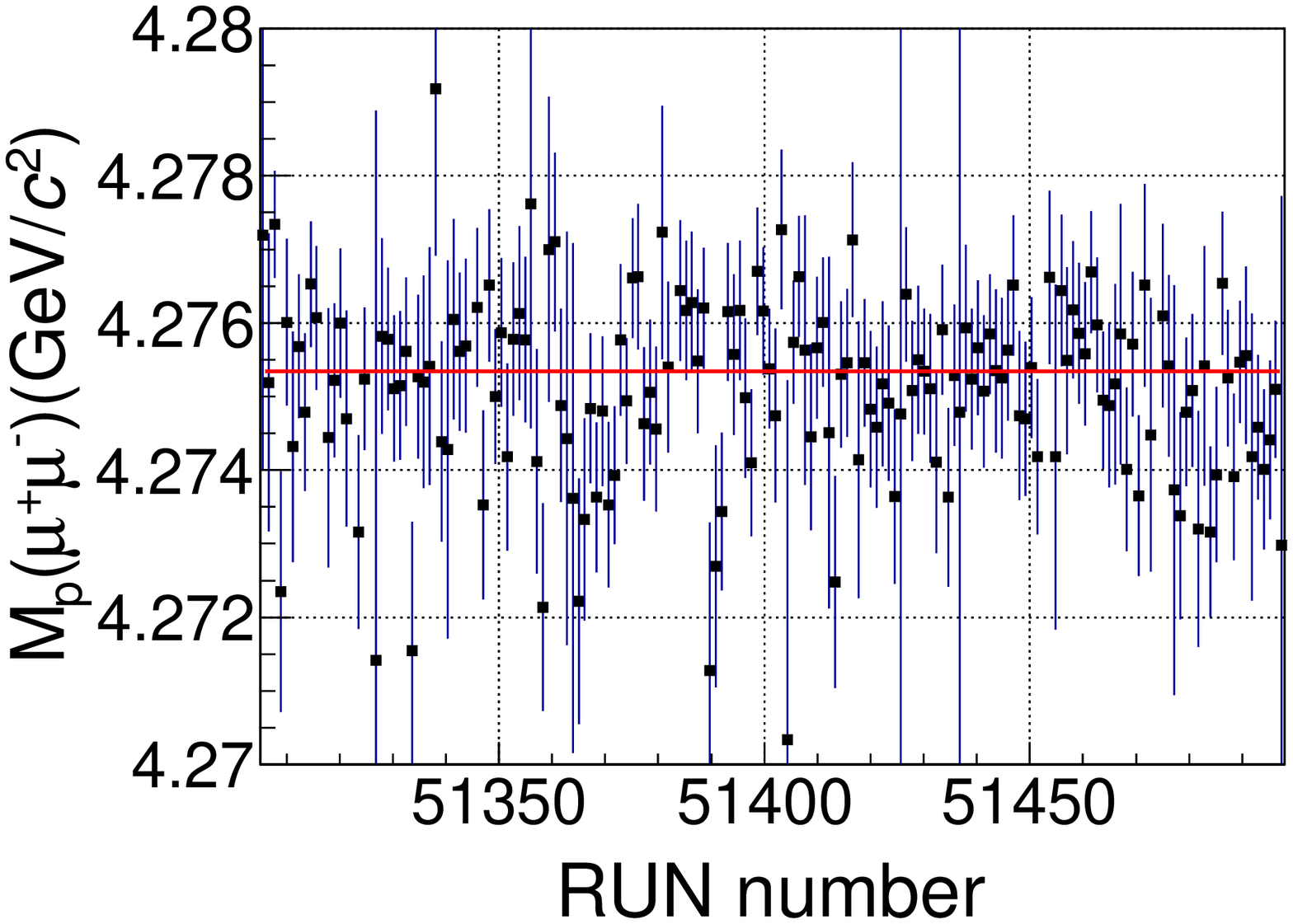}
  \includegraphics[width=0.22\textwidth]{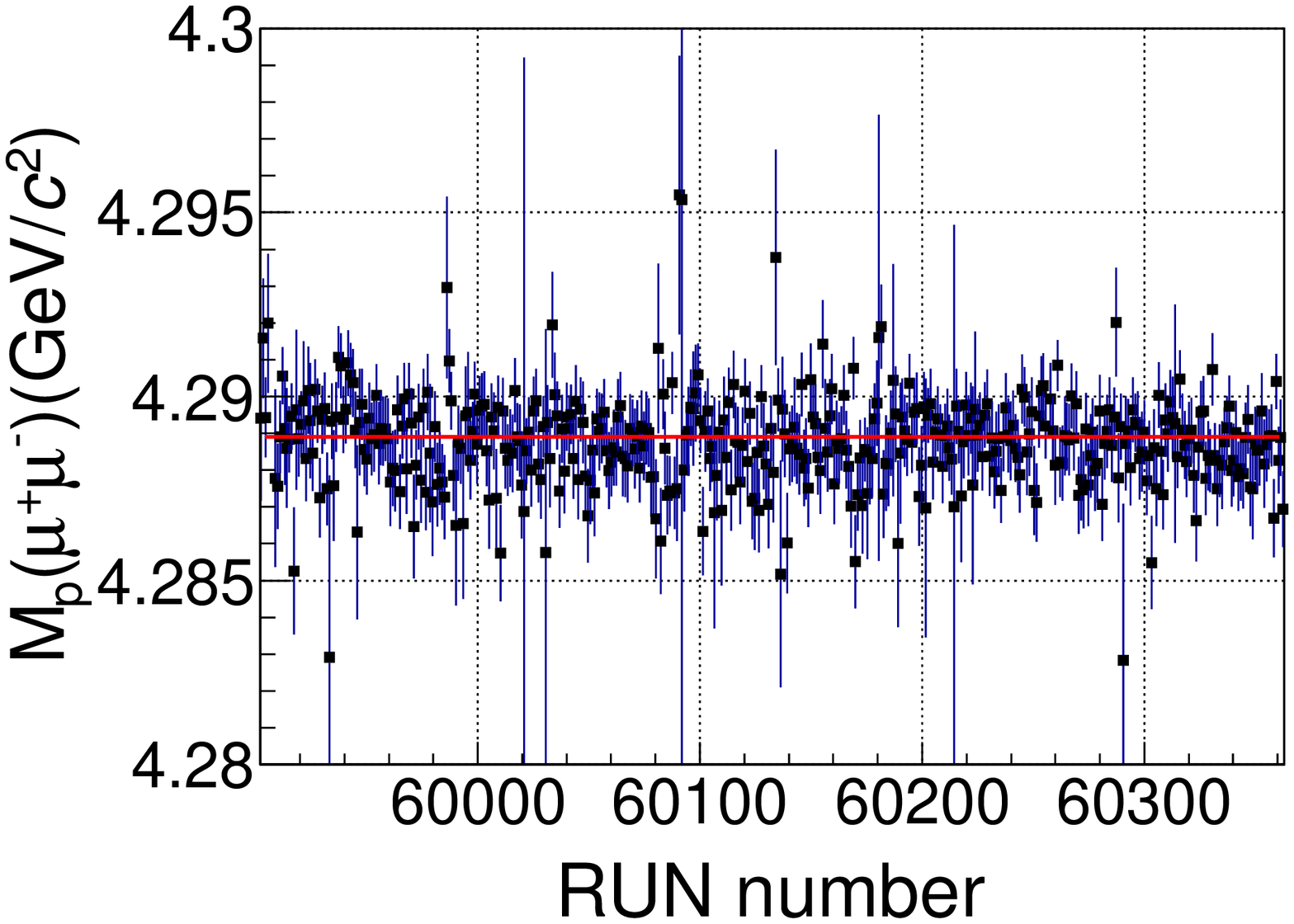}
  \includegraphics[width=0.22\textwidth]{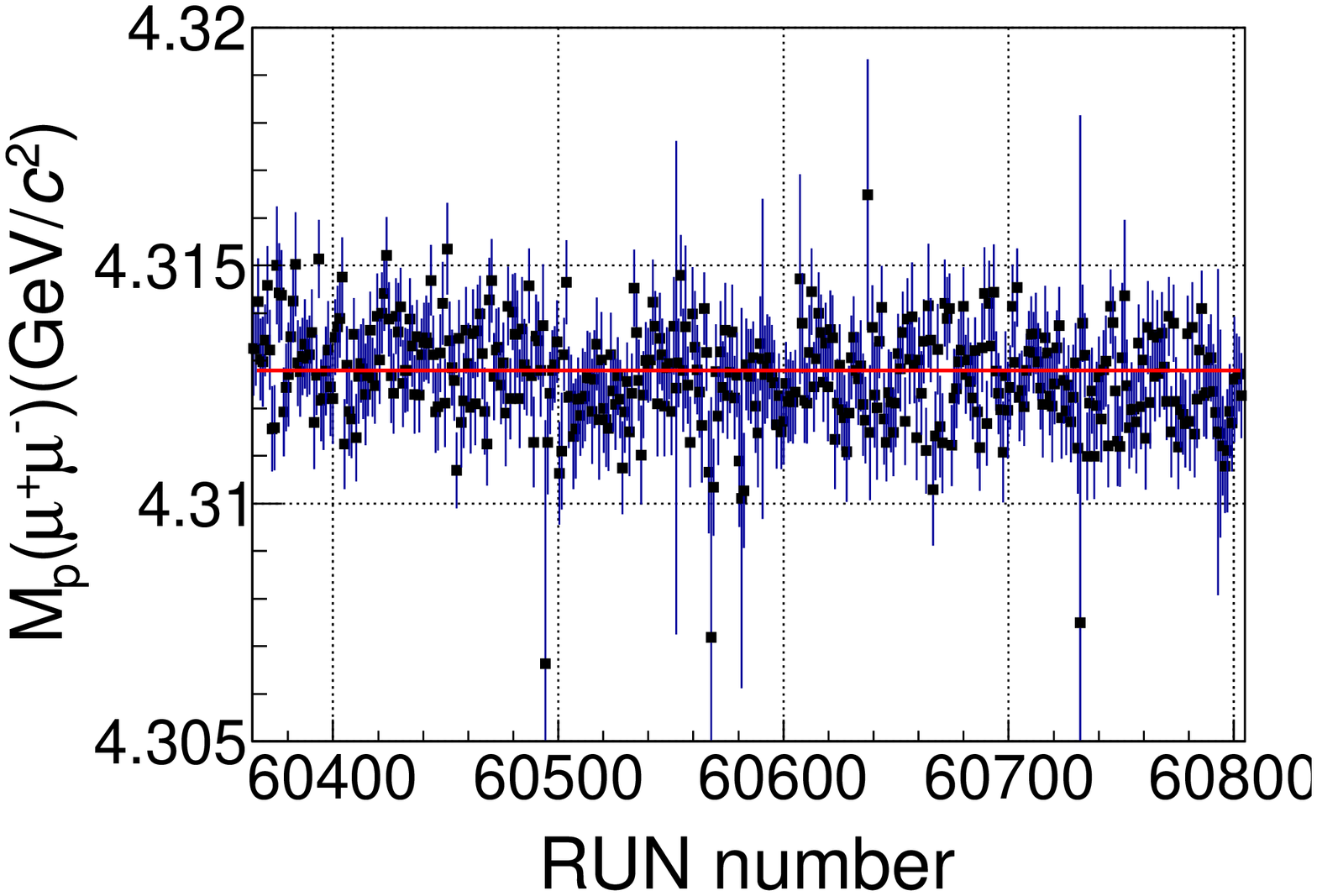}
  \includegraphics[width=0.22\textwidth]{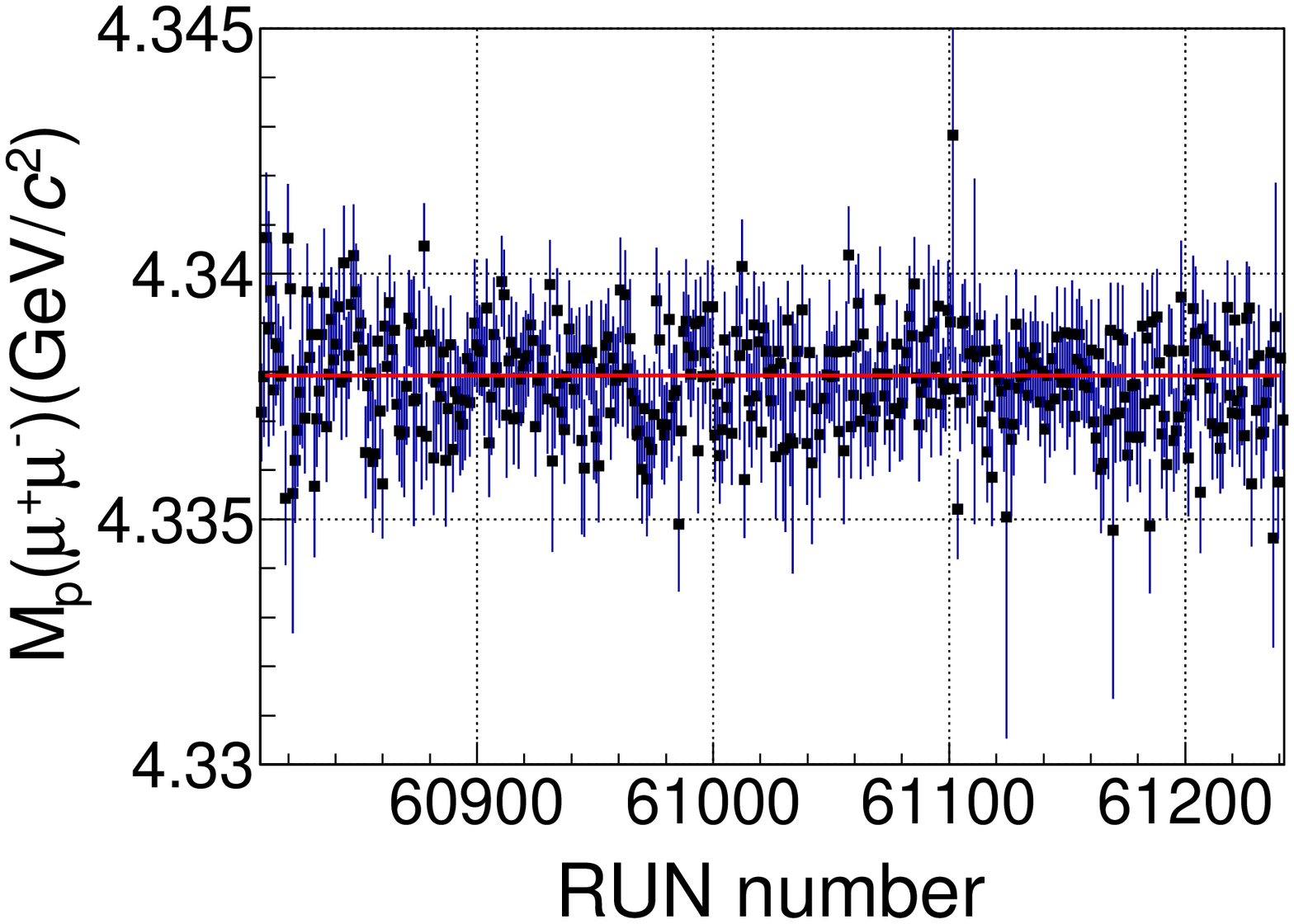}
  \includegraphics[width=0.22\textwidth]{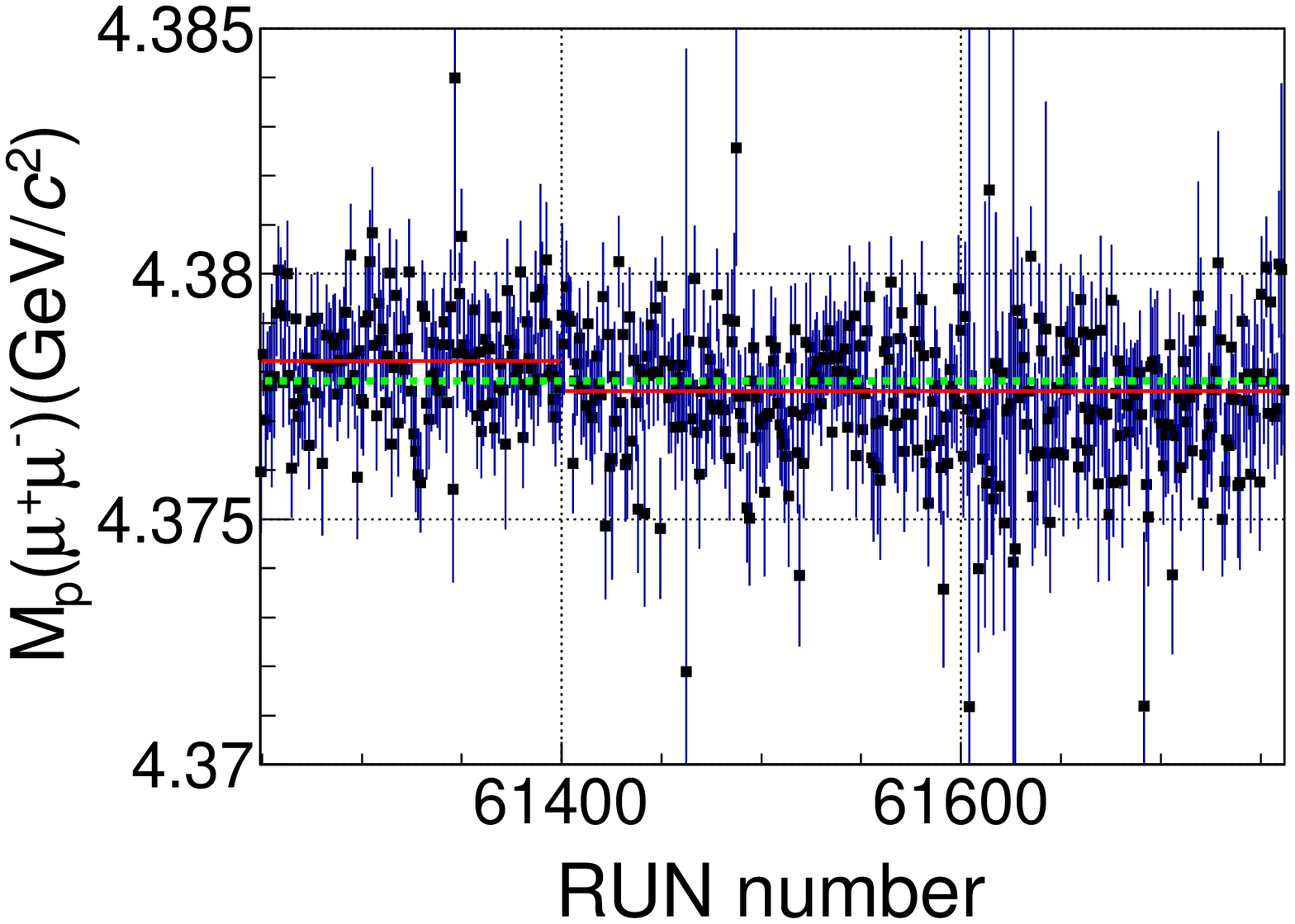}
  \includegraphics[width=0.22\textwidth]{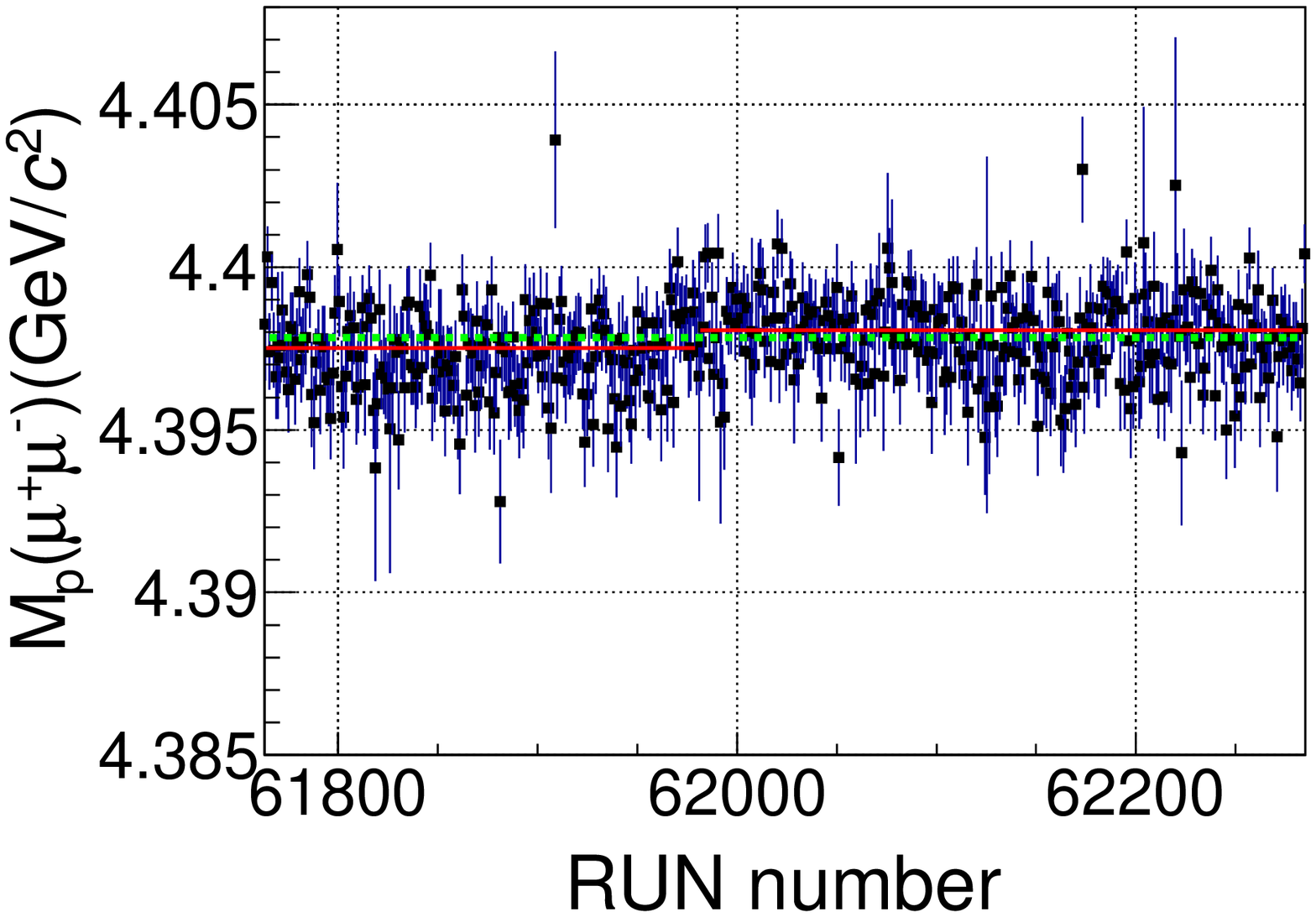}
  \includegraphics[width=0.22\textwidth]{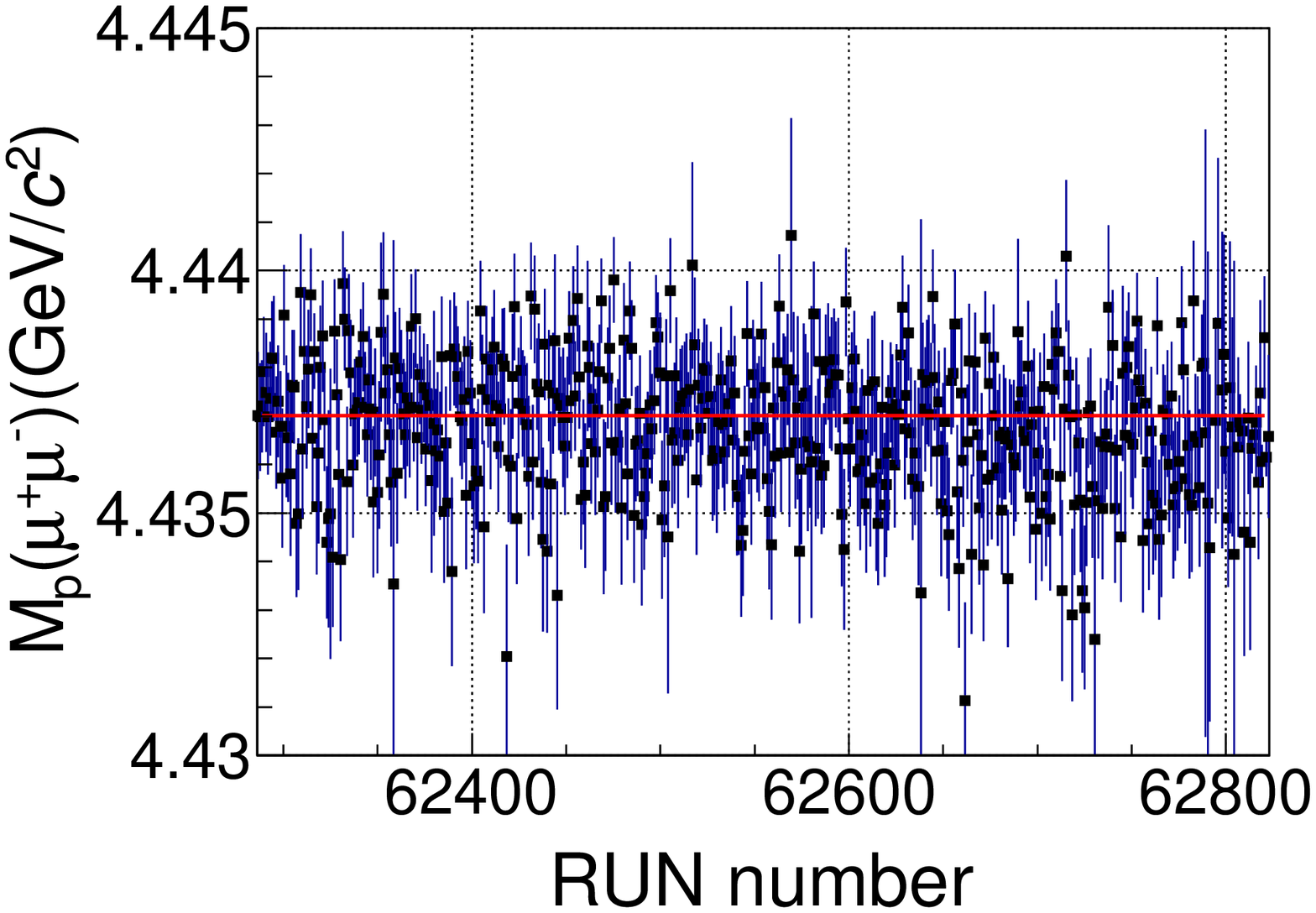}
\caption{Measured run-by-run values for the $M_{\rm p}(\MM)$ of di-muon events in
each data sample. The red solid lines show the fit results for the
data samples of each stable period of time. The green dotted lines
are the fit results of the entire sample when there is an energy
jump.} \label{rbrcheck}
\end{figure}

To examine the stability of the $\Ecm$ over the data-taking period
 for each data sample, the fit procedure is repeated for each
run of the data sample. The measured peak values of the $\MM$ invariant
mass distribution versus run number for all 16 samples are
shown in Fig.~\ref{rbrcheck}. There are small jumps of less than
1~MeV in the 4130, 4200, 4210, 4246, 4380, and 4400 samples. Before
and after the jumps, the energy is stable. We fit each stable part
of the distribution with a linear function and Table~\ref{ave}
summarizes the average, $M^{\rm ave}(\MM)$, for each period of time.
The deviation of $M^{\rm ave}(\MM)$ from the peak position
obtained in the full data sample is taken as one source of
systematic uncertainty.

\begin{table}[htbp]
 \centering
\caption{Average value of $M^{\rm ave}(\MM)$ (in MeV/$c^{2}$) for
each stable data-taking period within each data sample.}\label{ave}
\begin{tabular}{lllll}
\hline \hline
  Sample  &    Run Number     &   $M^{\rm ave}(\MM)$     &   Run Number  &   $M^{\rm ave}(\MM)$       \\
\hline
  4130    &   59163-59190     &   $4131.44\pm0.36$        &   59191-59573 &   $4130.02\pm0.05$        \\
  4160    &   59574-59896     &                           &               &   $4158.49\pm0.05$        \\
  4190    &   47543-48170     &                           &               &   $4187.52\pm0.06$        \\
  4200    &   48172-48290     &   $4197.14\pm0.12$        &   48291-48713 &   $4198.07\pm0.06$        \\
  4210    &   48174-49065     &   $4206.75\pm0.06$        &   49066-49239 &   $4207.49\pm0.09$        \\
  4220    &   49270-49787     &                           &               &   $4216.33\pm0.05$        \\
  4237    &   49788-50254     &                           &               &   $4233.21\pm0.04$        \\
  4246    &   50255-50520     &   $4241.01\pm0.08$        &   50521-50793 &   $4241.55\pm0.05$        \\
  4270    &   50796-51302     &                           &               &   $4265.20\pm0.06$        \\
  4280    &   51305-51498     &                           &               &   $4275.34\pm0.09$        \\
  4290    &   59902-60363     &                           &               &   $4288.91\pm0.05$        \\
  4315    &   60364-60805     &                           &               &   $4312.79\pm0.04$        \\
  4340    &   60808-61242     &                           &               &   $4337.93\pm0.05$        \\
  4380    &   61249-61400     &   $4378.23\pm0.09$        &   61401-61762 &   $4377.61\pm0.06$        \\
  4400    &   61763-61980     &   $4397.51\pm0.08$        &   61981-62285 &   $4398.06\pm0.07$        \\
  4440    &   62286-62823     &                           &               &   $4437.01\pm0.05$        \\
\hline \hline
\end{tabular}
\end{table}

\section{\boldmath Momentum calibration with ISR $J/\psi$ signal}

\begin{figure}[htbp]
\centering
  \subfigure[]{\includegraphics[width=0.45\textwidth, height=0.30\textwidth]{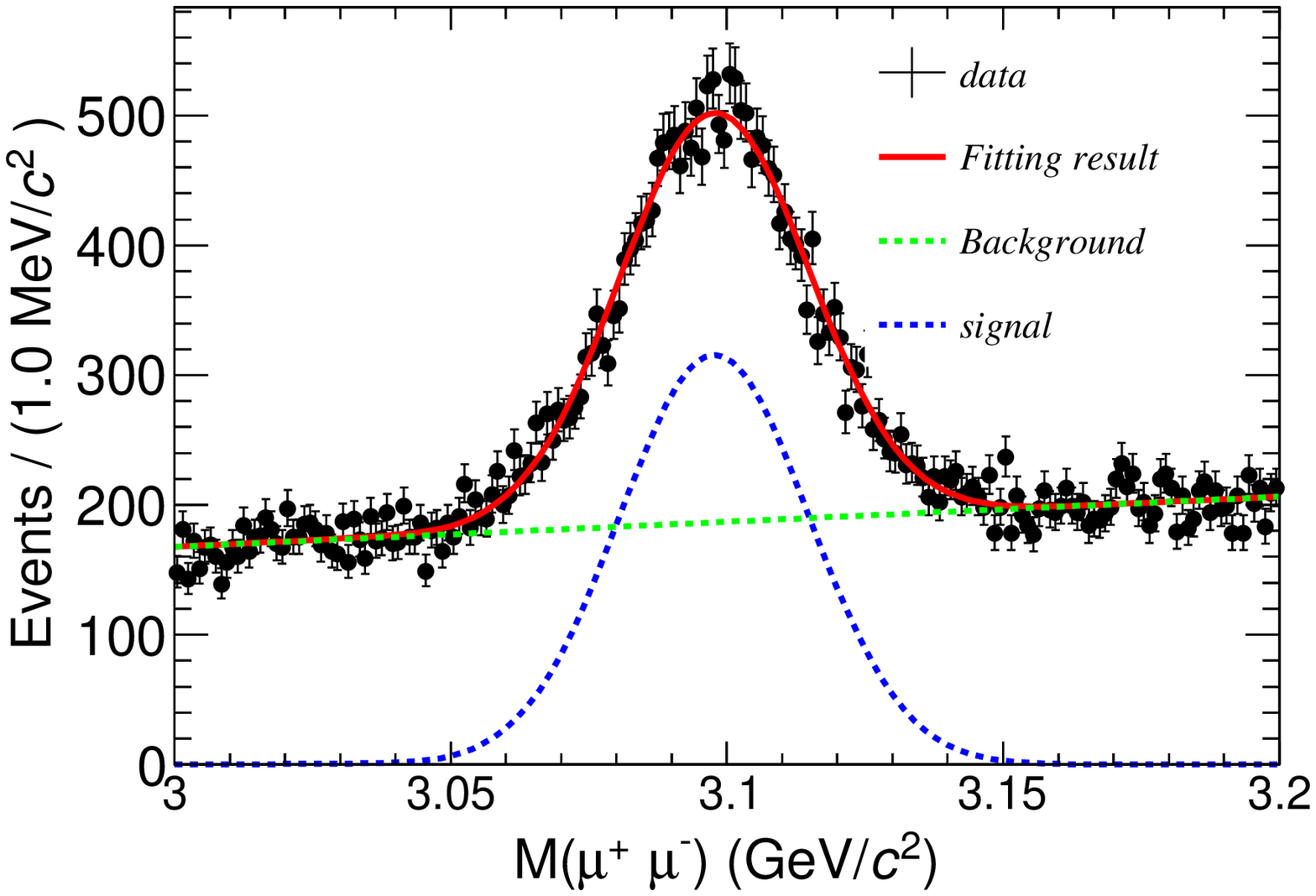}}
  \subfigure[]{\includegraphics[width=0.45\textwidth, height=0.30\textwidth]{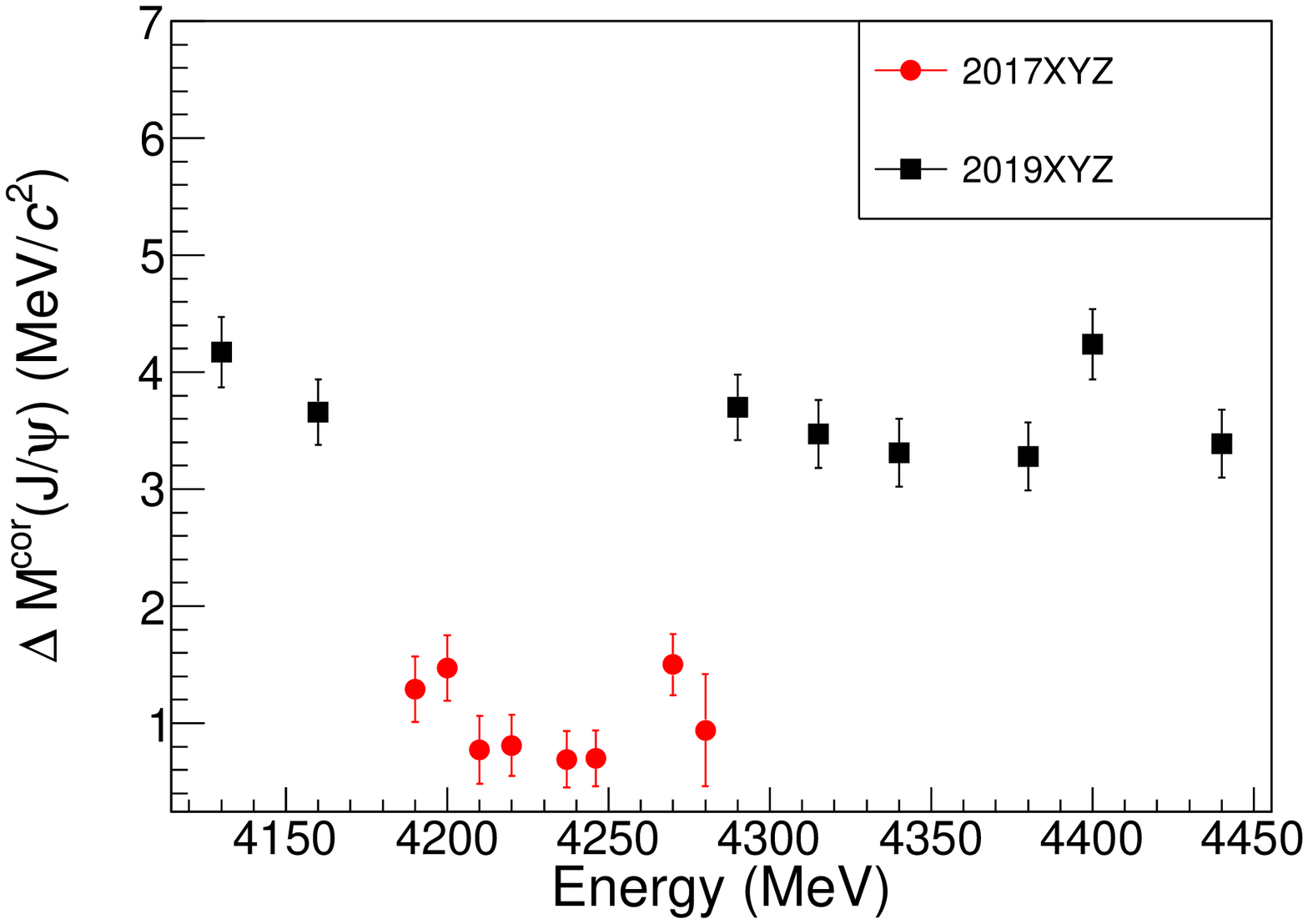}}
\caption{(a) Fit to the $M(\mu^{+}\mu^{-})$ distribution in the
$\jpsi$ signal region for the 4190 data sample. Black dots with
error bars are data, the red curve shows the fit result, the blue curve is for signal, and the green dashed line is for
background. (b) The difference between $M^{\rm cor}(J/\psi)$ and
the world average mass of $J/\psi$~\cite{PDG}, $\Delta M^{\rm cor}(J/\psi)$ for
each data sample.} \label{MJpsi_fit}
\end{figure}

The momentum measurement of the muon tracks is validated with
$\jpsi \to \MM$ candidates produced via the process $\EE\to
\gamma_{\rm ISR}\jpsi$ selected in the previous section. The
distribution of $M(\MM)$ of each sample is fitted with a
crystal-ball function~\cite{CB_func} for the $J/\psi$ signal and a
linear function to model the background from continuum production
of $\EE\to \gamma\MM$. Figure~\ref{MJpsi_fit}(a) shows the fit
result for the 4190 data sample as an example. The peak position
of the $\jpsi$ signal, $M^{\rm obs}(J/\psi)$, is used to calibrate
the momentum measurement of the muon tracks.

Due to FSR, $\jpsi\to \MM\gamma_{\rm FSR}$, the measured $M^{\rm
obs}(J/\psi)$ is slightly lower than the world average $J/\psi$
mass ($m_{\jpsi}$) given by the PDG~\cite{PDG}. The mass shift due to the FSR photon(s) $\Delta M^{\gamma J/\psi}_{\rm FSR}$ of the process $\EE\to \gamma_{\rm ISR}J/\psi$ at each $\Ecm$ is obtained by using the generator
{\sc photos}~\cite{Photos} with FSR turned on or off. The shift is around
0.3~MeV/$c^2$ with little dependence on the CM energy of the data
sample.

Comparing the $M^{\rm cor}(J/\psi)=M^{\rm obs}(J/\psi)+\Delta
M^{\gamma J/\psi}_{\rm FSR}$ (as shown in Table~\ref{ResultXYZ})
with the world-average $J/\psi$
mass value $m_{\jpsi}$ in Particle Data Book (PDG), we measure the bias in the $\jpsi$ mass measurement
($\Delta M^{\rm cor}(J/\psi)$) due to the muon track momentum
calibration, as shown in Fig.~\ref{MJpsi_fit}(b). It can be seen
that the bias in $\jpsi$ invariant mass is stable throughout one
running year, but is quite different in the 2017XYZ and 2019XYZ
samples. This may indicate that the calibrations in these two
periods of time have significant differences.

\begin{figure}[htbp]
\centering
  \includegraphics[width=0.45\textwidth]{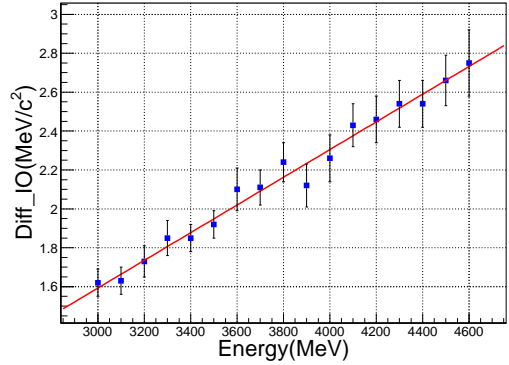}
\caption{The ($E_{\rm cm}^{\rm out} - E_{\rm cm}^{\rm in}$) are the difference between output and input $\Ecm$ (the output $\Ecm$ is equal to the $M_{\rm p}(\MM)$ if the events without radiation.) at different CM energies and simulated by di-muon events without radiation. The figures shows the dependence of the bias in $M_{\rm p}(\MM)$ 
due to the bias in track momentum calibration with di-muon events and it should be the same for data and MC (slope $k$).
} \label{Recdiff}
\end{figure}

Through MC simulation we find that the bias in $M(\MM)$
measurement depends on ($M(\MM)-m_{\jpsi}$) linearly (see
Fig.~\ref{Recdiff}), so the correction to the $M(\MM)$ due to
calibration is expressed as
\begin{equation}
\label{RecCorrection}
  \Delta M_{\rm cal} = -(k \times (\Ecm - m_{\jpsi}) + \Delta M^{\rm cor}(J/\psi)) (\rm{MeV}),
\end{equation}
where the slopes $k = (7.11\pm 0.50)\times 10^{-4}$ and $(7.04\pm
0.57)\times 10^{-4}$ are for the 2017XYZ and 2019XYZ samples,
respectively. They agree within the statistical uncertainties in
the MC samples, which indicates that the momentum dependence of the
calibration constants is very similar in the 2017XYZ and 2019XYZ
samples.

\section{\boldmath The mass shift $\Delta M_{\rm ISR/FSR}$}\label{Misfrcorrection}

The $\Ecm$ of the initial $\EE$ pair is measured via the di-muon
process $\EE\to (\gamma_{\rm ISR/FSR})\MM$. However, due to the
emission of radiative photons, the invariant mass of the $\MM$
pair is smaller than $\Ecm$ by $\Delta M_{\rm ISR/FSR}$. This
correction is estimated with MC simulation using {\sc
babayaga3.5}~\cite{babayaga}.

\begin{figure}[htbp]
  \centering
  \includegraphics[width=0.6\textwidth, height=0.40\textwidth]{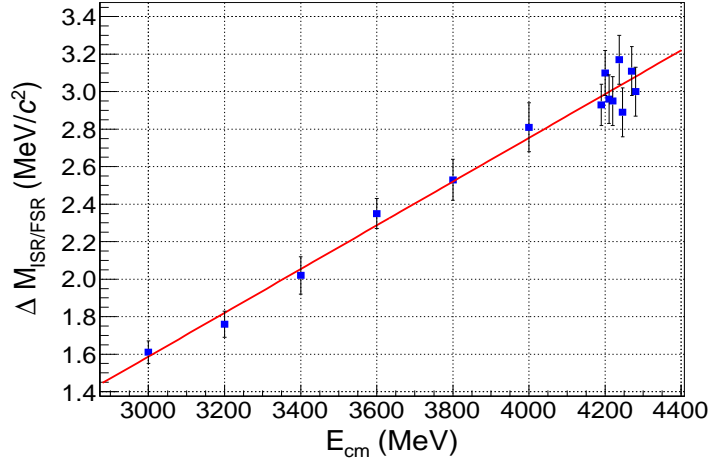}
\caption{The mass shift $\Delta M_{\rm ISR/FSR}$ versus CM energy
for $\EE\to (\gamma_{\rm ISR/FSR})\MM$ MC samples. The red solid
line is the fit with a linear function.} \label{radfit}
\end{figure}

We generate one million di-muon events for each sample with
ISR/FSR turned on or off, apply the same event selection criteria
to the di-muon events in data (as described in
Sec.~\ref{MeasEcms}), and fit the distributions of $M(\MM)$ from
the samples with ISR/FSR on and off with a Gaussian function in a
range around the peak (same as for data in Sec.~\ref{MeasEcms}).
The difference in $M_{\rm p}(\MM)$ is taken as the mass shift
$\Delta M_{\rm ISR/FSR}$ caused by ISR or FSR. The $\Delta M_{\rm ISR/FSR}$ versus $\Ecm$ shown in Fig.~\ref{radfit} indicates that
the ISR/FSR effect depends on $\Ecm$ linearly. The data are
fitted with a linear function to have an improved precision
measurement of the correction. The fit gives
\begin{equation}
\label{radcorrect}
 \Delta M_{\rm ISR/FSR} = ( 1.17\pm 0.05)\times 10^{-3}\times \Ecm
                        + (-1.91\pm 0.20) (\rm{MeV})
\end{equation}
with a correlation factor of $-0.99$ between the slope and the intercept,
and the goodness of the fit is $\chi^2/ndf = 6.2/12$.

\section{Systematic uncertainties}\label{Sys_err}

The systematic uncertainty in $\Ecm$ is from the momentum
calibration of $\mu^{\pm}$, the estimation of the mass shift
$\Delta M_{\rm ISR/FSR}$ due to ISR/FSR, the open angle cut of
$\cos\theta_{\MM}$, the corresponding fit procedure, and the
generator. The bias of the momentum measurement of $\mu^{\pm}$ and the
estimation of the mass shift $\Delta M_{\rm ISR/FSR}$ due to
ISR/FSR both have a linear relationship with $\Ecm$, and the uncertainty
caused by the uncertainty of the parameters is regarded as the
systematic uncertainties.

In order to reduce the influence of the events with large
radiation, we have required $\cos\theta_{\MM}<-0.9997$.
Different cut values will give different $M_{\rm p}(\MM)$ and
corresponding radiation correction values $\Delta M_{\rm
ISR/FSR}$. The changes in these two parts cancel each other out.
The biggest difference comes from the data between $-0.9997$ and
$-0.99975$, and is $0.12\pm 0.02$~MeV. We take
0.14~MeV as the uncertainty due to this requirement.

The $M_{\rm p}(\MM)$ is measured by fitting with a Gaussian
function in the range of $(-1\sigma,~+1.5\sigma)$ around the peak
with fit quality $\chi^2/ndf < 2.0$. If the fit range is smaller
than the standard range, the difference in fit results is less
than 0.1~MeV. We take this as the uncertainty due to the fit
method.

The contribution to the systematic uncertainty of the ISR/FSR
correction from the generator is negligibly small, as claimed in
Ref.~\cite{babayaga}. The uncertainties from other sources, such
as background and other event selection criteria, are negligible.

Assuming all the sources of systematic uncertainty are
independent, the total systematic uncertainty is obtained by
adding all the items in quadrature, which is listed in
Table~\ref{ResultXYZ}. The uncertainty is smaller than 0.6~MeV for
all the data samples.

\section{Summary}

The center-of-mass energies, $\Ecm$, of the data samples are
obtained by using Eq.~(\ref{Ecmscal}), with the correction factors
in Eqs.~(\ref{RecCorrection})~and~(\ref{radcorrect}). The final
results are listed in Table~\ref{ResultXYZ} including the
statistical and systematic uncertainties. The corresponding
statistical uncertainty is very small, and the systematic
uncertainty is less than 0.36 MeV everywhere, with
 the exception of the point at 4280MeV, where the error on $\Delta M^{\rm cor}$ is much larger than the rest.
 The stability of $\Ecm$ over time for the data samples is also examined.

The results presented in this work are essential for the discovery
of new states and the investigation of the transitions of charmonium
and charmoniumlike states~\cite{XYZ} using the BESIII data. Some
of the analyses have been presented in
Refs.~\cite{r1,r2,r3,r4,r5,r6}.

\acknowledgements

The BESIII collaboration thanks the staff of BEPCII and the IHEP computing center for their strong support. This work is supported in part by National Key Research and Development Program of China under Contracts Nos. 2020YFA0406300, 2020YFA0406400; National Natural Science Foundation of China (NSFC) under Contracts Nos. 11625523, 11635010, 11735014, 11822506, 11835012, 11935015, 11935016, 11935018, 11961141012; the Chinese Academy of Sciences (CAS) Large-Scale Scientific Facility Program; Joint Large-Scale Scientific Facility Funds of the NSFC and CAS under Contracts Nos. U1732263, U1832207; CAS Key Research Program of Frontier Sciences under Contracts Nos. QYZDJ-SSW-SLH003, QYZDJ-SSW-SLH040; 100 Talents Program of CAS; INPAC and Shanghai Key Laboratory for Particle Physics and Cosmology; ERC under Contract No. 758462; European Union Horizon 2020 research and innovation programme under Contract No. Marie Sklodowska-Curie grant agreement No 894790; German Research Foundation DFG under Contracts Nos. 443159800, Collaborative Research Center CRC 1044, FOR 2359, FOR 2359, GRK 214; Istituto Nazionale di Fisica Nucleare, Italy; Ministry of Development of Turkey under Contract No. DPT2006K-120470; National Science and Technology fund; Olle Engkvist Foundation under Contract No. 200-0605; STFC (United Kingdom); The Knut and Alice Wallenberg Foundation (Sweden) under Contract No. 2016.0157; The Royal Society, UK under Contracts Nos. DH140054, DH160214; The Swedish Research Council; U. S. Department of Energy under Contracts Nos. DE-FG02-05ER41374, DE-SC-0012069.

\end{document}